\documentclass[
aps,
nofootinbib,
superscriptaddress,
tightenlines,
notitlepage,
twocolumn,
showpacs,
floatfix,
]{revtex4-1}
\usepackage{amssymb,amsmath,bm,tensor,braket}
\usepackage[colorlinks]{hyperref}

\usepackage{graphicx}
\usepackage{dcolumn}
\usepackage{bm}
\usepackage{xcolor}

\newcommand{\Tabriz}{\affiliation{Faculty of Physics, University of Tabriz,
Tabriz 51666-16471, Iran}}
\newcommand{\BIT}{\affiliation{School of Physics, Beijing Institute of Technology, Beijing, 100081, China}}

\begin{document}

\preprint{APS/123-QED}

\title{Cosmological Study in Myrzakulov $F(R, T)$ Quasi-dilaton Massive Gravity}

\author{Sobhan Kazempour}\email{sobhan.kazempour1989@gmail.com}\BIT \Tabriz
\author{Amin Rezaei Akbarieh}\email{am.rezaei@tabrizu.ac.ir}\Tabriz
 

\begin{abstract}
This study explores the cosmological implications of the Myrzakulov $F(R, T)$ quasi-dilaton massive gravity theory, a modification of the de Rham-Gabadadze-Tolley (dRGT) massive gravity theory. Our analysis focuses on the self-accelerating solution of the background equations of motion, which are shown to exist in the theory. Notably, we find that the theory features an effective cosmological constant corresponding to the massive graviton, which has important implications for our understanding of the universe's accelerated expansion.
To assess the validity of the Myrzakulov $F(R, T)$ quasi-dilaton massive gravity theory, we employ two datasets: the Union2 Type Ia Supernovae (SNIa) dataset, consisting of 557 observations, and the Pantheon SNIa data, which includes 1048 SNe I-a events gathered from diverse SN I-a samples.
Our results demonstrate that the theory is capable of explaining the accelerated expansion of the universe without requiring the presence of dark energy. This finding supports the potential of the Myrzakulov $F(R, T)$ quasi-dilaton massive gravity theory as an alternative explanation for the observed cosmic acceleration.
Moreover, we investigate the properties of tensor perturbations within the framework of this theory and derive a novel expression for the dispersion relation of gravitational waves. Our analysis reveals interesting features of the modified dispersion relation in the Friedmann-Lema\^itre-Robertson-Walker (FLRW) cosmology, providing new insights into the nature of gravitational waves in the context of the Myrzakulov $F(R, T)$ quasi-dilaton massive gravity theory.
\end{abstract}


\maketitle


\section{Introduction}\label{sec:1}

It is well-known that the origin of late-time accelerated expansion of the Universe which is confirmed by several observational research is unknown and would be considered as the biggest puzzles in cosmology \cite{SupernovaSearchTeam:1998fmf,WMAP:2003elm,WMAP:2003xez,Caldwell:2003hz,SDSS:2005xqv,SDSS:2009ocz,SNLS:2011cra}. 
Although the general theory of relativity has a lot of successful predictions \cite{Will:2001mx,Will:2005va,Will:2014kxa}, it solely can explain the late-time accelerated expansion of the Universe by considering the cosmological constant or dark energy component. As the origin of dark energy is not clear and the cosmological constant has a problem \cite{Copeland:2006wr,Weinberg:1988cp,Peebles:2002gy}, there has been a tendency towards modified gravity theories throughout the years \cite{Carroll:2004de,Clifton:2011jh,Sotiriou:2008rp,Nojiri:2006ri,Bamba:2012cp,Nojiri:2010wj,Starobinsky:1980te,Nojiri:2005jg,Bahamonde:2015zma}.   

According to modern particle physics, general relativity is a unique theory of a massless Lorentz-invariant spin-2 particle \cite{Weinberg:1965rz}. Modifying gravity through the massive gravity theory offers a valuable approach to understanding the nature of gravity and its interactions. In this framework, a spin-2 massive graviton is thought to mediate the propagation of gravity, providing a fresh perspective on the fundamental laws of physics. Investigating the properties and behavior of such a graviton could lead to new insights into the structure of the universe and potentially reveal novel phenomena. It should be mentioned that a number of attempts have been made to describe graviton and its interactions to explain the current accelerated expansion of the Universe. In fact, the main effort is introducing a theory that would be stable and consistent with cosmological observations \cite{deRham:2014zqa,deRham:2010kj,Hinterbichler:2011tt,deRham:2010ik,Hassan:2011hr,Hassan:2011zd}.  

The introduction of a massive spin-2 field theory was first successfully achieved by Fierz and Pauli in 1939. They developed a Lorentz invariant linear theory that included consistent interaction terms, which were later interpreted as a graviton mass. In fact, its theory includes a specific combination of the mass terms to have five physical degrees of freedom \cite{Fierz:1939ix}. 
The theory of massive gravity has undergone a lot of changes over the years. In 1970, van Dam, Veltman, and Zakharov showed that the Fierz and Pauli theory could not reduce to general relativity (massless theory) in the limit of $m_{g}\longrightarrow 0$. This discontinuity in predictions is called vDVZ discontinuity \cite{Zakharov:1970cc,vanDam:1970vg}. In the following, Vainshtein presented a way to avoid vDVZ discontinuity. He proposed that one should consider the nonlinear completions of the Fierz-Pauli term instead of linear \cite{Vainshtein:1972sx}. On the other hand, Boulware and Deser claimed that there is a ghost instability in non-linear theory which is called the Boulware-Deser ghost \cite{Boulware:1972yco,Arkani-Hamed:2002bjr,Creminelli:2005qk}.
In line with these efforts to revive massive gravity theory, in 2010 de Rham, Gabadadze, and Tolley (dRGT) introduced a fully non-linear massive gravity theory without Boulware-Deser ghost. Their research proposes a novel approach to understanding the nature of gravity by introducing nonlinear interactions that could reveal the presence of a massive spin-2 field in a flat spacetime \cite{deRham:2010ik,deRham:2010kj}. However, this theory admits solely an open Friedmann-Lema\^itre-Robertson-Walker (FLRW) solution. In fact, there are not any stable solutions for a homogeneous and isotropic Universe \cite{DeFelice:2012mx,Gumrukcuoglu:2011zh}. Therefore, there are enough motivations to present new extension theories to find stable self-accelerating solutions in the context of massive gravity theory \cite{DAmico:2011eto,Gumrukcuoglu:2012aa,DAmico:2012hia,Mukohyama:2014rca,Huang:2012pe,DeFelice:2013tsa,Gumrukcuoglu:2013nza,Langlois:2014jba,deRham:2014naa,Kenna-Allison:2018izo,Kenna-Allison:2019tbu,Gumrukcuoglu:2020utx,Akbarieh:2021vhv,Aslmarand:2021qwn,Akbarieh:2022ovn,Kazempour:2022let,Kazempour:2022giy,Kazempour:2022xzy}.

In this work, we aim to introduce a novel extension of the non-linear dRGT massive gravity theory, Myrzakulov $F(R, T)$ quasi-dilation massive gravity theory, which endeavors to provide a fresh perspective on explaining the late-time accelerated expansion of the universe within the framework of FLRW cosmology. It is interesting to note that the quasi-dilaton part of the action introduces an extra scalar degree of freedom to the dRGT theory \cite{DAmico:2012hia}. But, because of instabilities, some extensions of quasi-dilaton massive gravity theory have been introduced \cite{Mukohyama:2014rca,DeFelice:2013tsa,Gumrukcuoglu:2013nza,Akbarieh:2021vhv}.

Modified gravity theories have been widely explored in recent years as an alternative to standard general relativity. One such theory is the $F(R, T)$ gravity, which generalizes the curvature-matter coupling by introducing a function $F(R, T)$ of the Ricci scalar $R$ and the trace of the energy-momentum tensor $T$. This theory was first introduced by Harko et al. \cite{Harko:2011kv} and has since been extended to various forms, including the Myrzakulov $F(R, T)$ gravity. The latter, introduced by Myrzakulov \cite{Myrzakulov:2012axz,Mueller-Hoissen:1983vya}, is a specific type of $F(R, T)$ gravity theory that exhibits a distinct behavior, where torsion can propagate independently of spin matter density, introducing a novel aspect to the theory. This unique feature sets Myrzakulov $F(R, T)$ gravity apart from other modified gravity theories and makes it an interesting candidate for further exploration.
Moreover, the Myrzakulov $F(R, T)$ gravity theory is an intriguing modification of traditional gravity theories, where $R$ represents the curvature scalar and $T$ denotes the torsion scalar. Notice that in this theory it could be possible to reproduce the unification of $F(R)$ and $F(T)$ gravity theories \cite{Myrzakulov:2012axz,Myrzakulov:2012qp}. We can point out a variety of extensions of the $F(R, T)$ gravity model which have been considered by several researchers \cite{Harko:2011kv,Godani:2018sbl,Debnath:2018wct,Tretyakov:2018yph,Sarkar:2022lir,Bertini:2023pmp}.

In this work, we aim to investigate the cosmological implications of combining Myrzakulov $F(R, T)$ gravity with dRGT massive gravity. By adding Myrzakulov $F(R, T)$ gravity to dRGT massive gravity, we introduce extra degrees of freedom to the obtained cosmological model, which can potentially provide the flexibility to avoid instability and strong coupling issues. Our goal is to find a suitable massive gravity model that can explain the late-time accelerated expansion of the Universe in FLRW cosmology, using the cosmological constant corresponding to the massive graviton term.

In order to constrain the parameters of the theory we consider the SNIa observational data. The theory compares with these data using the Bayesian statistic method which is on the basis of minimum $X^{2}$. Also, to ascertain the distributions of the parameters, we utilized the standard Bayesian technique, specifically employing the Markov Chain Monte Carlo (MCMC) method. Moreover, using the dispersion relation of gravitational waves one can impose the constraints on the theory. In addition, analysis of the dispersion relation of gravitation waves can be used in the phase evolution of the gravitational waveform.

The goals of the paper are to present a new extension of the massive gravity theory and find a stable self-accelerating solution to explain the late-time accelerated expansion of the Universe. Likewise, we try to present the constraints tools using cosmological data analysis and theoretical aspects of gravitational waves. The outline of this paper is organized as follows. In Sec. \ref{sec:2}, we present the new extension of massive gravity theory which is Myrzakulov $F(R, T)$ quasi-dilaton massive gravity theory. We obtain the background equations of motion and self-accelerating solution. In Sec. \ref{sec:3}, we constrain the parameters of the Myrzakulov $F(R, T)$ quasi-dilaton massive gravity theory using SNIa data by considering the Bayesian statistic technique and MCMC. In Sec. \ref{sec:4}, we undertake a perturbation analysis to derive the dispersion relationship of gravitational waves in the context of FLRW cosmology. In Sec. \ref{sec:5}, we indicate the conclusion.
In this paper, we consider natural units, where $c=\hbar=1$ and $M_{Pl}^{2}=8\pi G=1$, where $G$ is Newton's gravitational constant.

\section{$F(R, T)$ Quasi-dilaton Massive Gravity}\label{sec:2}

In this section, we present the new extension of the massive gravity theory. This theory is constructed by considering the Myrzakulov $F(R, T)$ modified gravity.
In the following, we show the cosmological background equations and self-accelerating solution.

The total action is:
\begin{eqnarray}
S=\frac{1}{2}\int d^{4} x \Bigg\lbrace \sqrt{-g}\bigg[ F(R, T) - \omega \partial_{\mu} \sigma \partial^{\mu} \sigma + 2{m}_{g}^{2} U(\mathcal{K}) \bigg] \Bigg\rbrace, \nonumber\\
\end{eqnarray}
where $\sqrt{-g}$ is the determinant of $g_{\mu\nu}$ which is the physical dynamical metric, $R$ is the curvature scalar, $T$ is the torsion scalar, $\omega$ is a dimensionless constant, $\sigma$ is the quasi-dilaton scalar and the last part of the action is related to the massive part. We consider the forms of the curvature and torsion scalars below \cite{Myrzakulov:2012qp,Myrzakulov:2012axz},
\begin{eqnarray}
F(R,T) = \xi R + \beta T,
\end{eqnarray}
where $\xi$ and $\beta$ are the constants. 
Here we should note that the $\sigma$ is the quasi-dilaton scalar and $\omega$ is a dimensionless constant of the quasi-dilaton theory \cite{DAmico:2012hia}. According to the Ref \cite{DAmico:2012hia}, the theory is invariant under a global dilation transformation, $\sigma \rightarrow \sigma +\sigma_{0}$. Furthermore, the value of $\omega$ should be in this range; $0 < \omega < 6$ in order to exist the stable self-accelerating background solutions \cite{DAmico:2012hia,Gumrukcuoglu:2013nza,Akbarieh:2021vhv}. 
However, according to the theory of gravity, the range value of $\omega$ can change, and this issue has something to do with the modified Friedman equations of the theories. 

It should be noted that the potential $U$ is the cause of the creation of the graviton mass $m_{g}$ which consists of three parts, i.e.
\begin{equation}\label{Upotential1}
U(\mathcal{K})=U_{2}+\alpha_{3}U_{3}+\alpha_{4}U_{4},
\end{equation}
with $\alpha_3$ and $\alpha_4$ dimensionless free parameters. Therefore, the potential sentences can be expressed as \cite{deRham:2010kj}
\begin{eqnarray}\label{Upotential2}
 U_{2}&=& \frac{1}{2} \big( [\mathcal{K}]^{2}-[\mathcal{K}^{2}]\big) ,
 \nonumber\\
 U_{3}&=& \frac{1}{6} \big( [\mathcal{K}]^{3}-3[\mathcal{K}][\mathcal{K}^{2}]+2[\mathcal{K}^{3}] \big) ,
 \nonumber\\
 U_{4}&=& \frac{1}{24} \big( [\mathcal{K}]^{4}-6[\mathcal{K}]^{2}[\mathcal{K}^{2}]+8[\mathcal{K}][\mathcal{K}^{3}]+3[\mathcal{K}^{2}]^2-6[\mathcal{K}^{4}]\big) , \nonumber\\
\end{eqnarray}
where "$[\cdot]$'' is interpreted as the trace of the tensor inside the brackets. Moreover, the building block tensor $\mathcal{K}$ is defined as
\begin{equation}\label{K}
\mathcal{K}^{\mu}_{~\nu} = \delta^{\mu}_{~\nu} - e^{\sigma}
\big(\sqrt{g^{-1}f}\big)^{\mu}_{~\nu},
\end{equation}
where $ f_{\alpha\nu}$ is the fiducial metric which is defined through
\begin{equation}\label{7}
f_{\alpha\nu}=\partial_{\alpha}\phi^{c}\partial_{\nu}\phi^{d}\eta_{cd},
\end{equation}
where $\eta_{cd}$ is the Minkowski metric ($c,d= 0,1,2,3$) and $\phi^{c}$ are the Stueckelberg fields which introduce to restore general covariance \cite{DAmico:2012hia,deRham:2010kj}.

\subsection{Background cosmological evolution}\label{subs21}
 
In this stage, we would like to consider the theory in an FLRW metric at the background level.  The dynamical metric, the vierbien, and the fiducial metric are expressed as
\begin{align}
\label{DMetric}
g_{\mu\nu}&={\rm diag} \left[-N^{2},a^2,a^2,a^2 \right],
\end{align}
\begin{align}
\label{DMetric}
e_{\mu}^{A}&={\rm diag} \left[N,a,a,a \right],
\end{align}
\begin{align}
\label{FMetric} 
f_{\mu\nu}&={\rm diag} \left[-\dot{f}(t)^{2},1,1,1 \right].
\end{align}
Also, $R$ and $T$ can be expressed as:
\begin{eqnarray}
R = g^{\mu\nu}R_{\mu\nu} + u, \nonumber\\
T= S_{\rho}^{~ \mu\nu} T^{\rho}_{~ \mu\nu} + v,
\end{eqnarray}
we consider the FLRW spacetime, and $u$ and $v$ can be defined,
\begin{eqnarray}\label{uv}
&& u =  6 \dot{j} + 18 \frac{\dot{a}}{aN} j + 6 j^{2} - 3 b^{2}, \nonumber\\ && v= 6 \big( j^{2} - b^{2} - (\frac{\dot{a}}{aN})^{2}\big),
\end{eqnarray}
also, we know 
\begin{eqnarray}
S_{\rho}^{~ \mu\nu} = \frac{1}{2} (C_{\rho}^{~\mu\nu}+\delta_{\rho}^{\mu}T_{\theta}^{~\theta\nu}-\delta_{\rho}^{\nu}T_{\theta}^{~\theta\mu}),
\end{eqnarray}
which $C$ is the contorsion \cite{Myrzakulov:2012qp}. In Eqs. (\ref{uv}), $b$ and $j$ are real functions and $N$ is a lapse function of the dynamical metric \cite{Myrzakulov:2012qp,Mueller-Hoissen:1983vya}. 
The geometrical roots of Myrzakulov $F(R, T)$ gravity are introduced elaborately \cite{Myrzakulov:2012qp}. The roots of $u$ and $v$ are related to the components of torsion and contorsion tensors.
In general, $u$ and $v$ can be defined as $u= u(t, a, \dot{a}, \ddot{a}, \dddot{a}, ...; f_{i})$ and $v= v(t, a, \dot{a}, \ddot{a}, \dddot{a}, ...; g_{i})$, while $f_{i}$ and $g_{i}$ are some unknown functions related with the geometry of the spacetime \cite{Myrzakulov:2012qp,Myrzakulov:2012axz,Mueller-Hoissen:1983vya,Myrzakulov:2012ug,Dimakis:1984jb,Gonner:1984rw,Capozziello:1994du,Tsamparlis:1981xm,Minkevich:1998cv,Capozziello:2001mq}.
Moreover, the orthonormal tetrad components $e_{i}(x^{\mu})$ are related to the metric as $g_{\mu\nu} = \eta_{ij} e_{\mu}^{i} e_{\nu}^{j}$. Note that in the paper, we only consider the above values of $u$ and $v$ which is the general form of Myrzakulov $F(R, T)$ gravity \cite{Myrzakulov:2012qp,Myrzakulov:2012axz,Myrzakulov:2012ug}. The Myrzakulov $F(R,T)$ gravity theory is rooted in Riemann-Cartan geometry, a framework that falls within the broader class of affine connected metric theories \cite{Saridakis:2019qwt,Conroy:2017yln}. This geometric structure is characterized by the presence of both non-zero curvature and torsion, and the action is formulated in terms of a general function of the Ricci scalar $R$ and the torsion scalar $T$.

It is obvious that $a$ is the scale factor and $\dot{a}$ is the derivative with respect to time.
In addition, $N$ is the lapse function of the dynamical metric, which relates the coordinate-time $dt$ to the proper-time $d\tau$ using $d\tau=Ndt$ \cite{Scheel:1994yr,Christodoulakis:2013xha}. Also, we express that the function $f(t)$ is the Stueckelberg scalar function, with $\phi^{0}=f(t)$ and $\frac{\partial\phi^{0}}{\partial t}=\dot{f}(t)$ \cite{Arkani-Hamed:2002bjr}.

Thus, we obtain the total pointlike Lagrangian in FLRW cosmology
\begin{eqnarray}
\mathcal{L}= \bigg[  \frac{-3 a\dot{a}^{2}\xi}{N} + \frac{3 a^{2}}{2}\bigg( a\big( 2 j^{2}(\beta + \xi) - b^{2} ( 2\beta +\xi ) \nonumber\\ + 2 \xi \dot{j} \big) N + 6 \xi j \dot{a}  \bigg)
\bigg]  + \frac{\omega a^{3}}{2N}\dot{\sigma}^{2} + m_{g}^{2}\Bigg\lbrace Na^{3}(Y \nonumber\\ - 1)
 \bigg[3(Y - 2)  - (Y - 4)(Y - 1)\alpha_{3}  - (Y \nonumber\\ - 1)^{2}\alpha_{4}\bigg]
+\dot{f}(t)a^{4}Y(Y - 1)\bigg[3 - 3(Y \nonumber\\ - 1)\alpha_{3} + (Y - 1)^{2}\alpha_{4}
\bigg] \Bigg\rbrace, \nonumber\\
\end{eqnarray}
where
\begin{eqnarray}\label{XX}
Y\equiv\frac{e^{\sigma}}{a}.
\end{eqnarray}
It is necessary to notice that the gauge transformations remove the unphysical fields from the theory on the classical level \cite{Grosse-Knetter:1992tbp}. Therefore, we regard the unitary gauge i.e., $f(t)=t$. So, a constraint equation would be achieved by varying with respect to $f$,
\begin{eqnarray}\label{Cons}
\frac{\delta \mathcal{L}}{\delta f} = m_{g}^{2} \frac{d}{dt} \lbrace 
a^{4}Y (Y - 1)  [3 - 3(Y - 1) \alpha_{3} \nonumber\\ + ( Y - 1 )^{2}\alpha_{4}]\rbrace =0.
\end{eqnarray}

We obtain the modified Friedman equation by considering the variation of the pointlike Lagrangian with respect to lapse function $N$,
\begin{eqnarray}\label{EqN}
\frac{1}{a^{3}}
\frac {\delta\mathcal{L}}{\delta N}= 3 H^{2}\xi - \frac{\omega}{2}\big( H+\frac{\dot{Y}}{YN} \big)^{2} + \frac{3}{2} \big( 2 j^{2} ( \beta +\xi ) \nonumber\\ - b^{2} ( 2\beta +\xi ) + 2 \xi \dot{j} \big) - m_{g}^{2}(Y-1)\bigg[(Y - 4)(Y \nonumber\\ - 1)\alpha_{3}
+(Y-1)^{2}\alpha_{4} - 3(Y-2)
\bigg]=0. \nonumber\\
\end{eqnarray}
The equation of motion related to the scale factor $a$ is given by
\begin{eqnarray}
\frac{1}{6 a^{2}N}\frac{\delta\mathcal{L}}{
\delta a}= \frac{1}{4}\bigg( 6 j^{2}\big( \xi +\beta \big) -3 b^{2}\big( 2\beta +\xi \big) + 6 \xi \nonumber\\ \big( H^{2} + \dot{j}^{2}\big)  \bigg) + \frac{\dot{H} \xi}{N} + \omega\big(\frac{\dot{\sigma}}{2N}\big)^{2} + \frac{m_{g}^{2}}{2} \bigg( 6 + 4 \alpha_{3} \nonumber\\ +\alpha_{4} - (2 + r ) ( 3 + 3\alpha_{3} + \alpha_{4})Y + ( 2r + 1) (1\nonumber\\+2\alpha_{3}+\alpha_{4})Y^{2}  - r (\alpha_{3} +\alpha_{4}) Y^{3} \bigg)=0. \nonumber\\
\end{eqnarray}
By varying the pointlike Lagrangian with respect to $\sigma$, the  equation of motion corresponding to the scalar field is obtained by
\begin{eqnarray}\label{EqSig}
\frac{1}{a^{3}N}\frac{\delta\mathcal{L}}{
\delta \sigma}= - 3 \frac{H^{2}\omega}{N} + m_{g}^{2}Y\Bigg\lbrace  6(r+1)\big(\alpha_{4}+2\alpha_{3} \nonumber\\ +1\big)Y -(3+r)\big(3+3\alpha_{3}+\alpha_{4}\big) - 3(3r+1)(\alpha_{4}\nonumber\\ +\alpha_{3}
)Y^{2}+4r\alpha_{4}Y^{3}\Bigg\rbrace =0, 
\end{eqnarray}
where $r\equiv\frac{a}{N}$ and  $H\equiv\frac{\dot{a}}{Na}$.
Note that the below equations are obtained by considering the notation in Eq (\ref{XX}). 
\begin{equation}
\frac{\dot{\sigma}}{N}= H+\frac{\dot{Y}}{NY}, \qquad \ddot{\sigma}=\frac{d}{dt}\Big(NH+\frac{\dot{Y}}{Y}\Big).
\end{equation}
According to the Stuckelberg field $f$ which introduces time reparametrization invariance, the Bianchi identity relates the four equations of motion. Therefore, the equation of motion has something to do with the scale factor is redundant and we disregard it.
\begin{eqnarray}
\frac{\delta S}{\delta \sigma}\dot{\sigma}+\frac{\delta S}{\delta f}\dot{f}-N\frac{d}{dt}\frac{\delta S}{\delta N}+\dot{a}\frac{\delta S}{\delta a}=0.
\end{eqnarray}

It should be pointed out that in particular conditions, all of the background equations and total Lagrangian reduce to those in Refs. \cite{DAmico:2012hia,Gumrukcuoglu:2013nza}.

\subsection{Self-accelerating background solutions}\label{subsec22}

In order to evaluate self-accelerating solutions in the context of this new extension, we integrate the Stueckelberg constraint equation (\ref {Cons}). Therefore, we have

\begin{eqnarray}\label{Self}
Y (Y-1)\bigg[3-3(Y-1)\alpha_{3}+(Y-1)^{2}\alpha_{4}\bigg] \propto a^{-4}. \nonumber\\
\end{eqnarray}
It is obvious that we look forward to explaining the accelerated expansion of the Universe in the context of Myrzakulov $F(R, T)$ quasi-dilaton massive gravity theory. The constant solution of the above equation leads to the effective energy density which behaves similarly to the cosmological constant. Therefore, in an expanding universe, the right-hand side of Eq. (\ref{Self}) decays as $a^{- 4}$, so the right-hand side of that equation would decrease in a sufficiently long time and makes the left-hand side of the equation equal to zero. Hence, $Y$ transforms into the saturate constant value $Y_{SA}$ which would be a root of the left-hand side of the equation (\ref{Self}).
\begin{equation}\label{Self2} 
(Y-1)\big[3-3(Y-1)\alpha_{3}+(Y-1)^{2}\alpha_{4}\big]\bigg|_{Y=Y_{\rm SA}}=0.
\end{equation}

Equation (\ref{Self2}) admits four distinct solutions. The one obvious solution is $Y=0$, however, this solution implies $\sigma\longrightarrow -\infty$. As this solution would be multiplied by the perturbations of the auxiliary scalars, we encounter strong coupling in the vector and scalar sectors. So, this solution has to disregard avoiding strong coupling \cite{DAmico:2012hia}. Furthermore, another solution that could be considered is $Y=1$. But, it must be eliminated because of a vanishing cosmological constant and inconsistency \cite{DAmico:2012hia}. As a result, there are solely two solutions which are given, 
\begin{equation}\label{XSa}
Y_{\rm SA}^{\pm}=\frac{3\alpha_{3}+2\alpha_{4}\pm\sqrt{9\alpha_{3}^{2}-12\alpha_{4}}}{2\alpha_{4}}.
\end{equation}
Using Eq. (\ref{EqN}) and Eq. (\ref{XSa}), we can obtain the modified Friedmann equation as below,
\begin{eqnarray}\label{EqFr}
 \bigg[ 3 H^{2} \xi + \frac{3}{2} \big( 2 j^{2} ( \beta +\xi ) - b^{2} ( 2\beta +\xi ) + 2 \xi \dot{j} \big) \nonumber\\ - \frac{\omega}{2}H^{2} \bigg] =\Lambda_{\rm SA}^{\pm}, \nonumber\\
\end{eqnarray}
where we have the effective cosmological constant corresponding to the massive part of the action
\begin{eqnarray}
\Lambda_{\rm SA}^{\pm}\equiv  m_{g}^{2}(Y_{\rm SA}^{\pm}- 1)\Big[ 
6-3Y_{\rm SA}^{\pm} +(Y_{\rm SA}^{\pm} - 4)(Y_{\rm SA}^{\pm} \nonumber\\  - 1)\alpha_{3}\nonumber +(Y_{\rm SA}^{\pm} - 1)^{2}\alpha_{4}\Big]. 
\end{eqnarray}
The above equation should be re-written by considering Eq. (\ref{XSa}),
\begin{eqnarray}
\Lambda_{\rm SA}^{\pm}=\frac{3m^{2}_{g}}{2\alpha^{3}_{4}}\bigg[9\alpha^{4}_{3}\pm 3\alpha^{3}_{3}\sqrt{9\alpha^{2}_{3}-12\alpha_{4}}-18\alpha^{2}_{3}\alpha_{4}\nonumber\\\mp 4\alpha_{3}\alpha_{4}\sqrt{9\alpha^{2}_{3}-12\alpha_{4}}+6\alpha^{2}_{4}\bigg].
\end{eqnarray}
Hence, $H^{2}$ in given by equation (\ref{EqFr}),
\begin{eqnarray}\label{H2}
H^{2}=\frac{- 6 j^{2} (\xi + \beta) + 3 b^{2} (\xi + 2\beta) -6 \xi \dot{j} + 2\Lambda_{\rm SA}^{\pm}}{6\xi - \omega}.
\end{eqnarray}
At the end of this subsection, we obtain $r_{\rm SA}$ using Eq. (\ref{EqSig}),
\begin{eqnarray}\label{rS1}
r_{\rm SA}=1+\frac{H^{2}\omega}{m_{g}^{2}Y_{\rm SA}^{2\pm}\big(\alpha_{3}Y_{\rm SA}^{\pm} - \alpha_{3} - 2\big)}.
\end{eqnarray}
Here, it should be expressed that we substituted $\alpha_{4}$ using the Stueckelberg equation (\ref{Self}). About the main result of this subsection, we presented that this theory consists of self-accelerating solutions with an effective cosmological constant i.e., $\Lambda_{\rm SA}^{\pm}$. Likewise, it is shown that we do have not any strong coupling, also the theory possesses a well-behaved self-accelerating solution that can explain the accelerated expansion of the Universe.
\begin{figure}
\centering
\includegraphics[width=7cm]{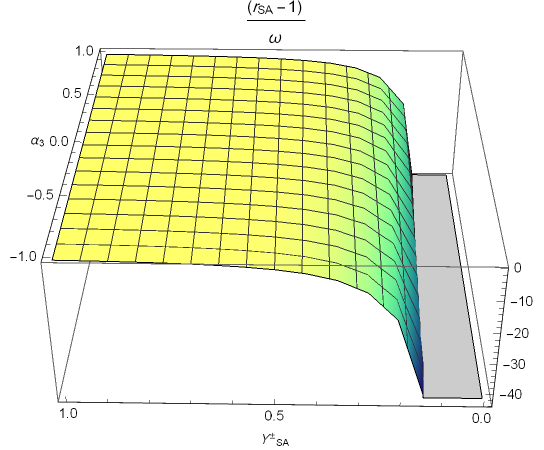}
\includegraphics[width=0.8cm]{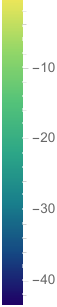}
\caption[figs]
{The expression $\frac{r_{\rm SA}-1}{\omega}$ can be obtained using Equation (\ref{rS1}), specifically when $m_{g}/H \simeq 1$. This simplification applies to the scenario in the case of $0 < Y_{\rm SA}^{\pm} < 1$. The regions that are excluded by this condition are depicted in gray.}
\label{fig1-1}
\end{figure}
\begin{figure}
\centering
\includegraphics[width=7cm]{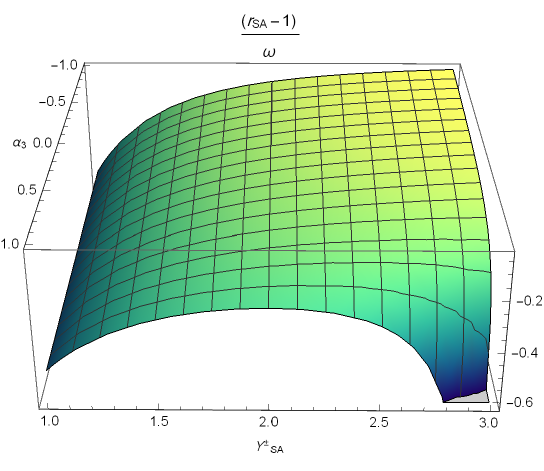}
\includegraphics[width=0.8cm]{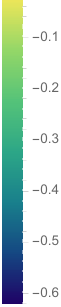}
\caption[figs]
{The expression $\frac{r_{\rm SA}-1}{\omega}$ can be obtained using Equation (\ref{rS1}), specifically when $m_{g}/H \simeq 1$. This simplification applies to the scenario in the case of $ Y_{\rm SA}^{\pm} > 1$. The regions that are excluded by this condition are depicted in gray.}
\label{fig2-2}
\end{figure}

Notice that we have illustrated limitations on the variable values of the theory. As a result, we have displayed the permitted parameter areas for equation (\ref{rS1}) in Figures (\ref{fig1-1}) and (\ref{fig2-2}). Bear in mind that these illustrations were crafted by assuming $m_{g}/H \simeq 1$ \cite{Kahniashvili:2014wua}. Furthermore, we wish to bring to attention that modifying the worth of the parameter $\alpha_{4}$ makes it conceivable to attain a sizeable value of $r_{\rm SA}$.

\section{Cosmological Data}\label{sec:3}

It is obvious that the exploration of type I-a supernovae showed the accelerated expansion of the Universe \cite{Copeland:2006wr,Frieman:2008sn,Perlmutter:2003kf,Yang:2019fjt}. In the following, using the Union2 supernovae I-a dataset consisting of 557 SNIa \cite{Amanullah:2010vv}, we would like to assess the Myrzakulov $F(R, T)$ quasi-dilaton massive gravity theory. Notice that the main goal is to constrain the theory using the cosmological data.
The results of the SNIa dataset should be represented in terms of $\mu_{\rm obs}$, and could be compared with the predictions of the model.

\begin{eqnarray}
\mu_{\rm th}(z_{i})=5\log_{10}D_{L}(z_{i})+\mu_{0},
\end{eqnarray}
note that $\mu_{0}=42.38 - 5 \log_{10} \mathcal{H}$ ($\mathcal{H}$ would be the Hubble constant $H_{0}$ in units of $100 \, {\rm km/s/Mpc}$), also we have the luminosity distance as below,
\begin{eqnarray}
D_{L}(z)=(1+z)\int_{0}^{z}\frac{dx}{E(x;p)},
\end{eqnarray}
moreover, it is clear that $E=H/H_{0}$ and $p$ represents the model parameters.
Notice that $X^{2}$ is given by
\begin{equation}\label{3.3}
X_{\mu}^{2}(p)=\sum_{i}\frac{[\mu_{\rm obs}(z_{i})-\mu_{\rm th}(z_{i})]^{2}}{\sigma^{2}(z_{i})},
\end{equation}
here $\sigma$ corresponds to $1\sigma$ error and the parameter $\mu_{0}$ is a nuisance parameter and is independent of the data points. We know that $X_{\mu}^{2}$ in equation (\ref{3.3}) make minimize it with respect to $\mu_{0}$ \cite{Nesseris:2005ur,DiPietro:2002cz}. 
\begin{equation}\label{3.4}
X_{\mu}^{2}(p)=\tilde{A}-2\mu_{0}\tilde{B}+\mu_{0}^{2}\tilde{C},
\end{equation}
where
\begin{eqnarray}
&&\tilde{A}(p)=\sum_{i}\frac{[\mu_{\rm obs}(z_{i})-\mu_{\rm th}(z_{i};\mu_{0}=0,p)]^{2}}{\sigma_{\mu_{\rm obs}}^{2}(z_{i})}, \nonumber\\
&&\tilde{B}(p)=\sum_{i}\frac{\mu_{\rm obs}(z_{i})-\mu_{\rm th}(z_{i};\mu_{0}=0,p)}{\sigma_{\mu_{\rm obs}}^{2}(z_{i})}, \nonumber\\
&&\tilde{C}=\sum_{i}\frac{1}{\sigma_{\mu_{\rm obs}}^{2}(z_{i})}.
\end{eqnarray}
It should be explained that for $\mu_{0}=\frac{\tilde{B}}{\tilde{C}}$, equation (\ref{3.4}) has a minimum at
\begin{equation}
\tilde{X}_{\mu}^{2}(p)=\tilde{A}(p)-\frac{\tilde{B}^{2}(p)}{\tilde{C}}.
\end{equation}
As it is clear that $X_{\mu, {\rm min }}^{2}=\tilde{X}_{\mu, {\rm min}}^{2}$, it could be considered minimizing $\tilde{X}_{\mu}^{2}$ which is independent of $\mu_{0}$. We should pay attention that the best-fit model parameters are determined by minimizing $X^{2}=\tilde{X}_{\mu}^{2}$. Meanwhile, we know that the corresponding $\mathcal{H}$ should be determined by $\mu_{0}=\frac{\tilde{B}}{\tilde{C}}$ for the best-fit parameters.

By considering equation (\ref{H2}), the change of variables $a=\frac{1}{1+z}$ and $\frac{d}{dt}=-H(z+1)\frac{d}{dz}$, the dimensionless Hubble parameter for this case should be written. Thus, the solution to the asymptotic state in small red-shifts should be given as,
\begin{eqnarray}
E=\frac{H(z)}{H_{0}}= 1 - \frac{\mathcal{M}z}{2(\mathcal{M}+\mathcal{D})}+ \frac{\big( 3\mathcal{M}^{2} +4\mathcal{M}\mathcal{D} \big) z^{2}}{8 (\mathcal{M}+\mathcal{D})^{2}} +....., \nonumber\\
\end{eqnarray}
where
\begin{eqnarray}
\mathcal{M} = \frac{-6 j^{2}(\xi +\beta) - 6\xi \dot{j}}{6\xi - \omega}, \nonumber\\
\mathcal{D} = \frac{3 b^{2}(\xi + 2\beta) + 2\Lambda_{\rm SA}^{\pm} }{6\xi -\omega}.
\end{eqnarray}

\subsection{Union2 I-a}\label{subsec31}

In this step, we plotted the $X^{2}$ and likelihoods as functions of parameters $\mathcal{M}$ and $\mathcal{D}$. According to calculations, the best fit has $X_{\rm min}^{2}=546.947$, and the best-fit parameters are,
\begin{align}
\mathcal{M}=& - 0.037, \\
\mathcal{D}=& 0.070.
\end{align}
Consequently, it should be mentioned the best fit for the Myrzakulov $F(R, T)$ quasi-dilaton massive gravity are shown using $\mathcal{H}=0.693$.

\begin{figure}
\centering
\includegraphics[width=6.5cm]{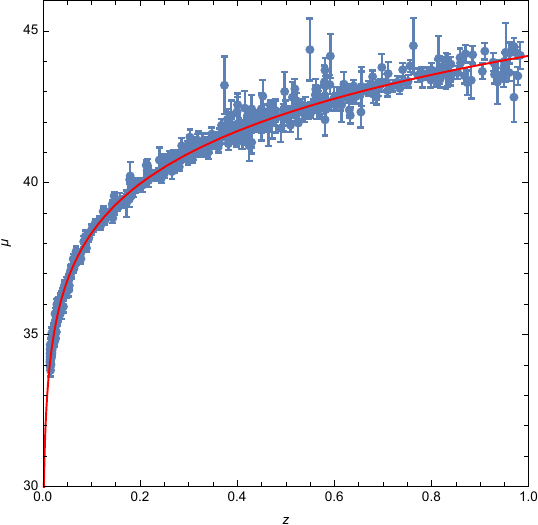}
\caption[figs]
{The distance modulus diagram for the best fit (red solid line) of the parameters of the Myrzakulov $F(R, T)$ quasi-dilaton massive gravity theory in comparison with the 557 Union2 SNIa data points (blue dots).}
\label{fig1}
\end{figure}

\begin{figure}
\centering
\includegraphics[width=6.5cm]{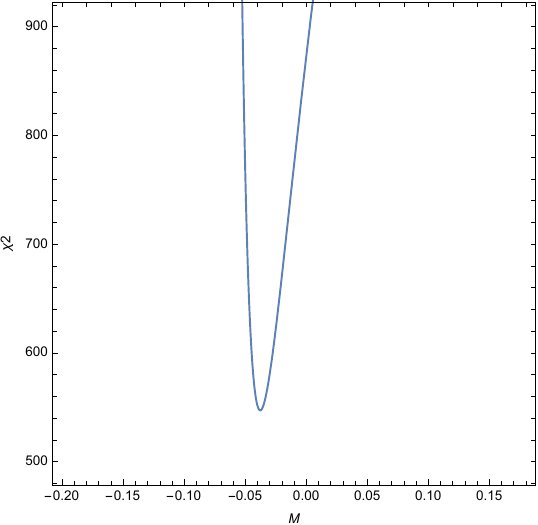}
\caption[figs]
{The $X^{2}$ functions of parameter $\mathcal{M}$ for the Myrzakulov $F(R, T)$ quasi-dilaton massive gravity.}
\label{fig2}
\end{figure}

\begin{figure}
\centering
\includegraphics[width=6.5cm]{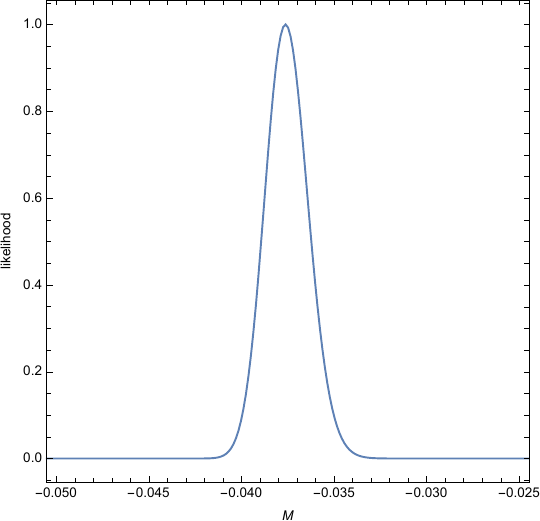}
\caption[figs]
{The likelihood as functions of parameter $\mathcal{M}$ for the Myrzakulov $F(R, T)$ quasi-dilaton massive gravity.}
\label{fig3}
\end{figure}

\begin{figure}
\centering
\includegraphics[width=6.5cm]{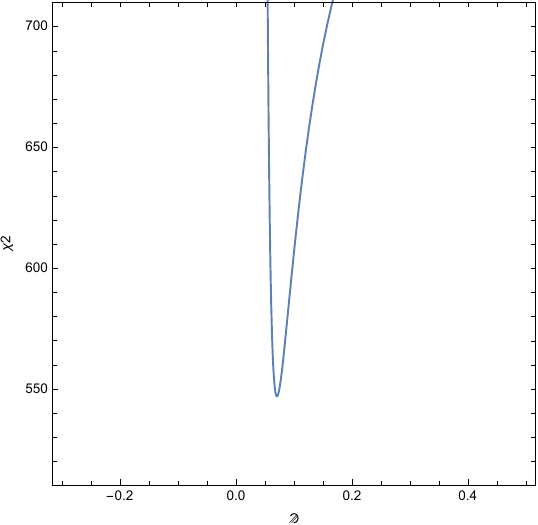}
\caption[figs]
{The $X^{2}$ functions of parameter $\mathcal{D}$ for the Myrzakulov $F(R, T)$ quasi-dilaton massive gravity.}
\label{fig4}
\end{figure}

\begin{figure}
\centering
\includegraphics[width=6.5cm]{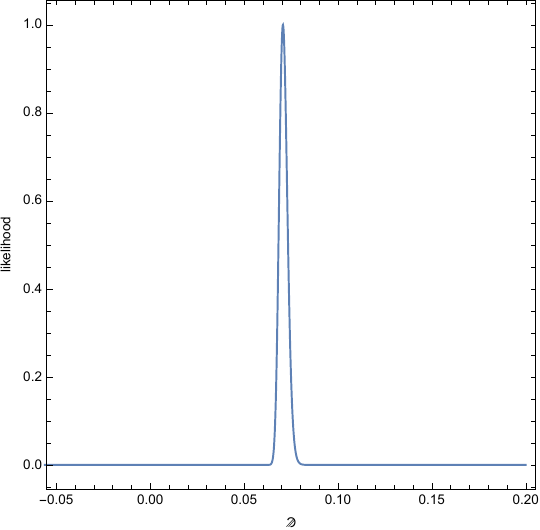}
\caption[figs]
{The likelihood as functions of parameter $\mathcal{D}$ for the Myrzakulov $F(R, T)$ quasi-dilaton massive gravity.}
\label{fig5}
\end{figure}
According to Figs (\ref{fig1}-\ref{fig5}), the result of fitting the Myrzakulov $F(R, T)$ quasi-dilaton massive gravity theory with the cosmological data gives us the best values of model parameters. 

\begin{figure}
\centering
\includegraphics[width=8.5cm]{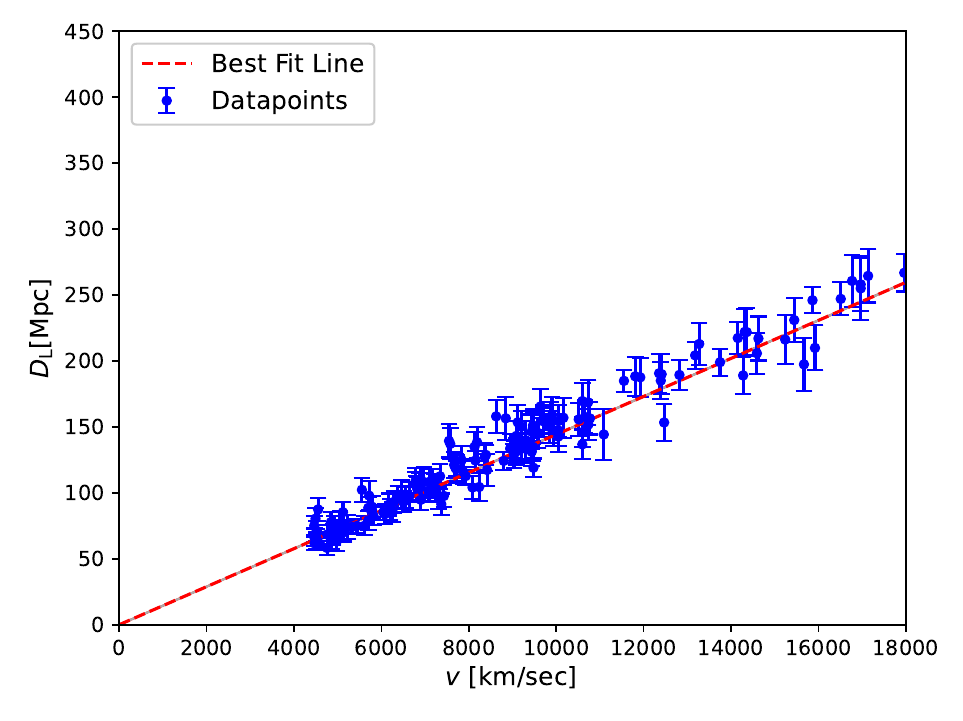}
\caption[figs]
{The best-fit diagram for the luminosity distance $D_{L}$ with respect to velocity $V$ for the Myrzakulov $F(R, T)$ quasi-dilaton massive gravity.}
\label{fig-51}
\end{figure}
In Fig (\ref{fig-51}), we plotted the velocity $V$ versus luminosity distance $D_{L}$ for the Myrzakulov $F(R, T)$ quasi-dilaton massive gravity theory. For calculating the velocity from redshift, one should use $V = c - \frac{2 c}{(z+1)^{2}+1}$.  

To test the viability of the Myrzakulov $F(R, T)$ quasi-dilaton massive gravity theory, we utilize the Union2 supernovae I-a dataset consisting of 557 SNIa. We employ standard frequentist approach to constrain the model parameters using the SNIa data. Specifically, we plot the $X^{2}$ and likelihood functions as functions of the parameters $\mathcal{M}$ and $\mathcal{D}$, and determine the best-fit values of these parameters by minimizing $X^{2}$.

\subsection{Pantheon}\label{subsec32}

During this stage, we utilized the Pantheon supernovae I-a data, encompassing 1048 SNe I-a events sourced from various SN I-a samples \cite{Pan-STARRS1:2017jku}. To determine the distributions of the parameters, we employed the standard Bayesian technique, specifically utilizing the MCMC method.
The resulting best-fit parameters are as follows:
\begin{eqnarray}
\mathcal{M} = - 0.150, \nonumber\\ \mathcal{D}= 0.155 .
\end{eqnarray}

\begin{figure}[h]
\centering
\includegraphics[width=7cm]{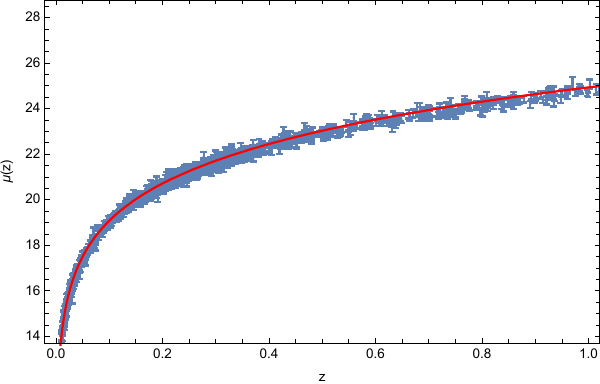}
\caption[The distance modulus diagram for the best fit (red solid line) of the parameters of the Myrzakulov $F(R, T)$ quasi-dilaton massive gravity]
{The distance modulus diagram for the best fit (red solid line) of the parameters of the Myrzakulov $F(R, T)$ quasi-dilaton massive gravity, compared with the Pantheon data points (blue dots) with error bar.}\label{4MCMC1}
\end{figure}
\begin{figure}
\centering
\includegraphics[width=7cm]{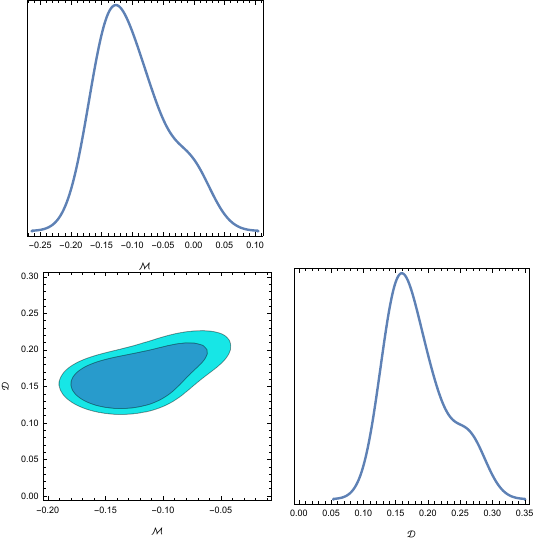}
\caption[The best values of the parameters of the Myrzakulov $F(R, T)$ quasi-dilaton massive gravity]
{The best values of $\mathcal{M}$ and $\mathcal{D}$ for the Myrzakulov $F(R, T)$ quasi-dilaton massive gravity.}\label{4MCMC2}
\end{figure}
\begin{figure}
\centering
\includegraphics[width=6cm]{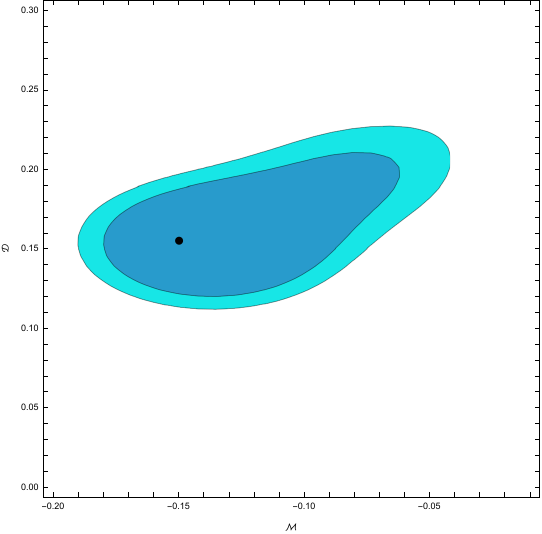}
\caption[The best values for $\mathcal{M}$ and $\mathcal{D}$ from the MCMC analysis.]
{The best values for $\mathcal{M}$ and $\mathcal{D}$ which has been shown by a black dot. The black point shows the best values from the MCMC analysis.}\label{4MCMC3}
\end{figure}

Our results provide strong evidence in favor of the Myrzakulov $F(R, T)$ quasi-dilaton massive gravity theory as a viable explanation for the accelerated expansion of the Universe. In fact, these results exhibit good agreement between the theoretical predictions and observational data. These findings lend support to the idea that the Myrzakulov $F(R, T)$ quasi-dilaton massive gravity theory can successfully describe late-time cosmic acceleration.

The variation in the best-fit values of the parameters $\mathcal{M}$ and $\mathcal{D}$ arising from the Union2 I-a and Pantheon data sets can be attributed to a multitude of factors, offering a nuanced insight into the complexities of cosmological analysis. Firstly, the sheer volume and diversity of the Pantheon data set, boasting 1048 SNe I-a events, surpasses that of Union2 with its 557 data points. This increased size enhances the robustness of parameter constraints and can subtly shift the best-fit values. Moreover, the Pantheon data set's broader range of SNe I-a samples contributes to the divergence in results.

The quality and precision of the data also play a pivotal role. The Pantheon data set is distinguished by its advanced data reduction techniques and meticulous calibration, enhancing the accuracy of the results. Differences in systematic and statistical uncertainties between the two data sets further influence the inferred parameter values.

Methodological disparities also come into play. The Pantheon analysis employs the MCMC method within a Bayesian framework, offering a more exhaustive exploration of the parameter space, whereas a simpler $X^{2}$ minimization approach is used for Union2. This distinction can lead to variations in the results.

Additionally, the inherent assumptions and complexities of the Myrzakulov $F(R, T)$ quasi-dilaton massive gravity theory itself influence the outcome. The choice of priors, initial conditions, and specific model variants can sway the inferred parameter values, with different data sets favoring distinct regions of the parameter space.

Lastly, we must consider the role of statistical fluctuations. Given the intricate nature of cosmological data and the presence of inherent uncertainties, it is natural for variations to arise between different data sets. The specific treatment of nuisance parameters and outliers can further accentuate these fluctuations.

Therefore, the disparity in the best-fit values of $\mathcal{M}$ and $\mathcal{D}$ stems from a combination of factors, encompassing data set size and diversity, data quality, methodological nuances, model assumptions, and the inherent statistical fluctuations inherent to cosmological investigations. The enhanced volume and improved data reduction techniques of the Pantheon data set are particularly influential in shaping the observed differences in parameter values.

The derived values of the model parameters and the Hubble constant provide further credence to this theory. Future investigations involving larger and more diverse datasets will help to reinforce or challenge these conclusions, ultimately contributing to a deeper understanding of the fundamental laws governing the behavior of the Universe.

We used a statistical technique called MCMC to narrow down the values of the model parameters in the Myrzakulov $F(R,T)$ quasi-dilaton massive gravity theory. We focused on two parameters, $\mathcal{M}$ and $\mathcal{D}$, which are connected to the model's underlying parameters. By including the $\Lambda_{\rm SA}^{\pm}$ parameter in our analysis, we were able to avoid a challenge in determining the values of these underlying parameters.
Our analysis involved running a large number of simulations (100,000 steps) to explore the possible values of the model parameters. We used a specific algorithm to ensure that the simulations were efficient and accurate. We also checked to make sure that the simulations were not correlated with each other, and that they had converged to a stable solution. The results of the analysis provided us with the best-fit values for the model parameters and their associated uncertainties.

\section{Perturbations Analysis}\label{sec:4}

In this section, we demonstrate the perturbation analysis of Myrzakulov $F(R, T)$ quasi-dilaton massive gravity.
It is worth noting that the significance of such analysis is that the stability conditions of the solutions can be determined.
As we would be interested in quadratic perturbations, we should expand the physical metric $g_{\mu\nu}$, the vierbein and the scalar field in terms of small fluctuations $\delta g_{\mu\nu}$, $\delta e^{A}_{\mu}$ and $\delta\sigma$ around  the background solution $g_{\mu\nu}^{(0)}$ and $\delta e^{A (0)}_{\mu}$:
\begin{equation}
g_{\mu\nu}=g_{\mu\nu}^{(0)}+\delta g_{\mu\nu},
\end{equation}
\begin{equation}
e^{A}_{\mu}=e^{A (0)}_{\mu} + \delta e^{A}_{\mu},
\end{equation}
\begin{equation}
\sigma = \sigma^{(0)} + \delta\sigma.
\end{equation}
Note that one can keep all terms up to quadratic order.
We consider tensor perturbations around background $\delta g_{ij}= a^{2}h_{ij}^{TT}$.
Also, the tensor perturbations are transverse $\partial^{i}h_{ij}=0$, 
and traceless $h_{i}^{~ i}=0$. 
We should pay attention that all perturbations are functions of time and space, and they are 
consistent with the transformations under spatial rotations 
\cite{Kahniashvili:2014wua,DeFelice:2013tsa}.

Besides, one can write the actions expanded in Fourier plane waves, namely
$\vec{\nabla}^{2}\rightarrow -k^{2}$, $d^{3}x\rightarrow d^{3}k$. Also, the spatial indices are raised and lowered by $\delta^{ij}$ and $\delta_{ij}$.
Hence, as all calculations are done in the unitary gauge, one does not need to specify gauge-invariant combinations \cite{Gumrukcuoglu:2013nza}.

\subsection{Dispersion Relation of GWs}

Notice that we obtain the perturbed action at the second order separately for the different parts. 

Before that, we present the dispersion relation of gravitational waves to the standard Einstein's theory below;
\begin{eqnarray}
S^{(2)}_{\rm Ein}=\frac{1}{8} \int d^{3}k dt a^{3}N  \bigg[ \frac{\dot{h}_{ij}\dot{h}^{ij}}{N^{2}} - \Big(\frac{k^{2}}{a^{2}} + M_{GW}^{2} \Big) h^{ij}h_{ij}\bigg], \nonumber\\ 
\end{eqnarray}
\begin{eqnarray}
M_{GW}^{2} = \frac{4\dot{H}}{N} + 6H^{2}. 
\end{eqnarray}
 
The Myrzakulov $F(R, T)$ part is given as
\begin{eqnarray}
S^{(2)}_{\rm F(R,T)}=\frac{1}{8} \int d^{3}k dt a^{3}N \Bigg\lbrace \xi \bigg[ \frac{\dot{h}_{ij}\dot{h}^{ij}}{N^{2}} - \Big(\frac{k^{2}}{a^{2}}+\frac{4\dot{H}
}{N} \nonumber\\+6H^{2} + u \Big) h^{ij}h_{ij}\bigg] + \beta \bigg[
\frac{\dot{h}_{ij}\dot{h}^{ij}}{N^{2}} -
\Big(\frac{k^{2}}{a^{2}} \nonumber\\ + 3H^{2} + v \Big)h^{ij}h_{ij} \bigg] \Bigg\rbrace. \nonumber\\
\end{eqnarray}
Meanwhile, the quasi-dilaton part of the perturbed action reads as
\begin{equation}
S^{(2)}_{\rm Quasi-dilaton}=-\frac{1}{8}\int d^{3}k dt a^{3}N\Bigg[
\left( \frac{\omega}{N^{2}} \dot{\sigma}^{2}\right) h^{ij}h_{ij}\Bigg],
\end{equation}

also, the massive gravity part becomes
\begin{eqnarray}
S^{(2)}_{\rm massive}= \frac{1}{8}\int d^{3}k dt a^{3}N 
m_{g}^{2}\bigg[(\alpha_{3}+\alpha_{4})rY^{3}\nonumber\\
-(1+2\alpha_{3}+\alpha_{4})(1+3r)Y^{2}
\nonumber\\ +(3+3\alpha_{3}+\alpha_{4})(3+2r)Y\nonumber\\
-2(6+4\alpha_{3}+\alpha_{4})\bigg]h^{ij}h_{ij}.
\end{eqnarray}

In the end, by considering the above terms, the second-order perturbed action for tensor perturbations $S^{(2)}_{\rm total}=S^{(2)}_{\rm F(R,T)}+S^{(2)}_{\rm Quasi-dilaton}+S^{(2)}_{\rm massive}$, becomes
\begin{eqnarray}\label{Sto}
S^{(2)}_{\rm total}=\frac{1}{8}\int d^{3}k \, dt \, a^{3}N\Bigg\lbrace 
\frac{\dot{h}^{ij}\dot{h}_{ij}}{N^{2}} (\xi + \beta) \nonumber\\
-\bigg[ \frac{k^{2}} {a^{2}} (\xi + \beta) +M_{\rm GW}^{2}\bigg]
h^{ij}h_{ij}\Bigg\rbrace , 
\end{eqnarray}
where
\begin{eqnarray}\label{M_GW}
M^{2}_{\rm GW}=\bigg(\frac{4\dot{H}}{N} + 6 H^{2} + u \bigg) \xi + \bigg( 3 H^{2} +v \bigg) \beta \nonumber\\ + \frac{\omega}{N^{2}} \dot{\sigma}^{2} +\varpi, \nonumber\\
\end{eqnarray}
and with
\begin{widetext}
\begin{eqnarray}
\varpi = \frac{m_{g}^{2}}{(Y_{\rm SA}^{\pm}-1)} \Bigg\lbrace Y_{\rm SA}^{\pm}\bigg[ 18 + 8 \alpha_{3} + Y_{\rm SA}^{\pm} \bigg( 2\alpha_{3}Y_{\rm SA}^{\pm} + Y_{\rm SA}^{\pm} - r_{\rm SA}\big( 3(\alpha_{3}+2) \nonumber\\+ Y_{\rm SA}^{\pm}(\alpha_{3}Y_{\rm SA}^{\pm} -4\alpha_{3}-3)\big) -8\alpha_{3}-10\bigg)\bigg] -2(\alpha_{3}+3) \Bigg\rbrace.\nonumber\\
\end{eqnarray}
\end{widetext}

The last relation is obtained using (\ref{XSa}) to substitute $\alpha_{3}$ and $\alpha_{4}$.
It is obvious the equation (\ref{M_GW}) determines the modified dispersion 
relation of gravitational waves in Myrzakulov $F(R, T)$ quasi-dilaton massive gravity.
In order to guarantee the stability of long-wavelength gravitational waves, the mass square of gravitational waves should be positive $M_{\rm GW}^{2}>0$ (i.e., Eq. (\ref{M_GW}) $>0$). But, if the mass square would be negative, it is unstable and can be considered as tachyonic. So, the instability could take the age of the Universe to develop by considering the mass of the tachyon in the order of the Hubble scale.

Notice that we showed the stability condition of the system, also the other main result of this section is the modified dispersion relation of gravitational waves. It is obvious that the achieved results in the paper is completely different from the standard Einstein's theory. They showed the propagation of gravitational perturbations in the FLRW cosmology in the Myrzakulov $F(R, T)$ quasi-dilaton massive gravity. It is worth mentioning that cosmological events such as gravitational wave observations can be used for testing modified propagation. The modification introduced in this theory will lead to additional effects on the phase evolution of the gravitational waveform. Moreover, these extra contributions could be detected with accurate matched-filtering techniques in the data analysis. It is interesting to note that Nishizawa and Arai investigated the modified propagation of gravitational waves in the cosmological context \cite{Nishizawa:2017nef,Arai:2017hxj,Nishizawa:2019rra}. The result of this paper also demonstrates theory-specific analysis in Myrzakulov $F(R, T)$ quasi-dilaton massive gravity which is complementary to their work.
At the end of this part, I should like to point out that future space-based gravitational-wave detectors which are sensitive, provide the testing of crucial properties of gravitation at different wavelengths.

\section{Conclusions}\label{sec:5}

In this study, we have successfully extended the dRGT massive gravity theory to develop the innovative Myrzakulov $F(R, T)$ quasi-dilaton massive gravity theory. Our exhaustive investigation has led to the derivation of the complete set of field equations for a FLRW background, paving the way for a detailed examination of the late-time acceleration of the Universe. Indeed, we have presented a comprehensive exposition of the new action and total Lagrangian, along with the complete set of equations of motion for a FLRW background. The exploration of extended massive gravity theories holds significant importance in unraveling the mysteries behind the late-time acceleration of the Universe. With this in mind, we have devoted a detailed discussion to elucidating the self-accelerating background solutions corresponding to the massive graviton. Furthermore, we have outlined an approach to interpreting the late-time acceleration of the universe within the framework of the Myrzakulov $F(R,T)$ quasi-dilaton massive gravity theory.

To validate the predictions of the Myrzakulov $F(R, T)$ quasi-dilaton massive gravity theory, we embarked on a rigorous evaluation by comparing its forecasts with the Union2 SNIa dataset, comprising 557 observations and the Pantheon SNIa encompassing 1048 SNe I-a events sourced from various SNIa samples. Our analysis revealed an exceptional agreement between the theoretical framework and the empirical data, strongly supporting the potential of the model in accurately describing the late-time cosmic acceleration. 

Motivated by the quest to deepen our understanding of gravitational phenomena, we conducted a comprehensive exploration of tensor perturbations. This endeavor allowed us to unveil the underlying principles governing the mass of the graviton within the context of the Myrzakulov $F(R, T)$ quasi-dilaton massive gravity theory. Our findings facilitated the derivation of the dispersion relation for gravitational waves and an examination of the propagation properties of gravitational perturbations in FLRW cosmology. These critical investigations hold significant implications for the current era of gravitational wave astronomy, where novel approaches to understanding the behavior of gravity are being actively pursued.

As we mentioned before, it is important to highlight that our work complements previous studies, such as those by Nishizawa and Arai, who investigated the modified propagation of gravitational waves in the cosmological context. Our research provides a theory-specific analysis in Myrzakulov $F(R, T)$ quasi-dilaton massive gravity, offering valuable insights into the nature of gravity in diverse environments.

As we conclude this paper, it is essential to acknowledge the profound impact that future space-based gravitational-wave detectors will have on our understanding of the universe. With their heightened sensitivity, they will enable the testing of crucial properties of gravitation at various wavelengths, further expanding our knowledge of the cosmos.

At the end, to provide a clearer understanding of the advantages of this new model, we have provided Table \ref{tab:table1} which shows the differences between our work and existing works.
\begin{table*}
\caption{\label{tab:table1} Features of cosmological models.}
\begin{tabular}{|c|c|c|c|c|c|c|c|c|}\hline
Cosmological Moles &Curvature&Torsion&Quasi-dilaton&Massive&Accelerating Solutions& Stability \\ \hline
Einstein-Hilbert & Yes & No & No & No & Cosmological Constant Problem \cite{Weinberg:1988cp} & -  \\ \hline
$F(R, T)$ & Yes & Yes & No & No & Unphysical solutions \cite{Amendola:2015ksp,Dolgov:2003px} & No \cite{Dolgov:2003px} \\ \hline
Myrzakulov $F(R, T)$ & Yes & Yes & No & No & Yes \cite{Myrzakulov:2012qp,Saridakis:2019qwt} & No \cite{Jamil:2011ptc}  \\ \hline
dRGT Massive Gravity & Yes & No & No & Yes &  Only Open Universe \cite{DeFelice:2012mx}& No \cite{DeFelice:2012mx,Gumrukcuoglu:2011zh}  \\ \hline
Quasi-dilaton Massive Gravity & Yes & No & Yes & Yes & Yes \cite{DAmico:2011eto} & No \cite{DeFelice:2013tsa,Mukohyama:2014rca} \\ \hline
Myrzakulov $F(R, T)$ Q-d M-G & Yes & Yes & Yes & Yes & Yes & Yes  \\ \hline
\end{tabular}
\end{table*}
\\

 \clearpage
\section*{Acknowledgements}
This work is based upon research funded by the University of Tabriz, Iran National Science Foundation (INSF), and Iran National Elites Foundation (INEF), under project No. 4014244. 
This work is supported by the National Key R$\&$D Program of China (grant 2023YFE0117200) and the National Natural Science Foundation of China (Nos. 12105013).

The authors are grateful to A. Emir Gumrukcuolu for useful comments. Also, the authors are grateful to Nishant Agarwal for notes and codes related to tensor perturbations.\\
 


\end{document}